# Response toward Public Health Policy Ambiguity and Insurance Decisions

Qiang Li

**Abstract:** Adjustments to public health policy are common. This paper investigates the impact of COVID-19 policy ambiguity on specific groups' insurance consumption. The results show that sensitive groups' willingness to pay (WTP) for insurance is 12.2% above the benchmark. Groups that have experienced income disruptions are more likely to suffer this. This paper offers fresh perspectives on the effects of pandemic control shifts.

**Qiang Li**

School of Finance Nankai University

No. 38 Tongyan Road, Jinnan District, Tianjin, P.R. China

Email: liqiang_dh@mail.nankai.edu.cn

# 1. Introduction

The study of insurance behavior is frequently associated with ambiguity and risk aversion (Hogarth and Kunreuther, 1989). Most public policies are characterized by ambiguity at the start of adjustment to compromise the interests of the actors and integrate opinions until a certain degree of consensus is reached. This paper examines the differences in insurance decisions of individuals with different responses to policy ambiguity using data from an online experiment. China's State Council issued steps to optimize epidemic prevention and control in November 2022[1], which were generally considered a forerunner to a substantial adjustment of COVID-19 control. When the number of confirmed cases surged, Shijiazhuang city removed the lockdown, reduced nucleic acid testing stations, and lowered the verification of nucleic acid results in public places. The series of moves by the Shijiazhuang authorities is unprecedented. Despite repeated statements by governments at all levels that the "dynamic zero COVID-19 strategy" would not alter, the fact is that public health policy implementation is ambiguous. Therefore, we studied how residents react to this ambiguity and whether they are more willing to pay for commercial health insurance.

We conducted our study in the context of ambiguity in the extraordinary COVID-19 response policies of the Chinese authorities, utilizing an online experiment with the inhabitants of Shijiazhuang. Our results demonstrate that after controlling for risk factors in the sample area, the group with positive self-action attitudes in the face of ambiguity policies (i.e., the sensitive group) increased their willingness to pay (WTP) for commercial health insurance by 12.2% in comparison to the benchmark. We also find that residents who faced income interruption as a result of pandemic control were more likely to pay for commercial insurance. These findings imply that public health policy ambiguity has a significant impact on certain groups' economic behavior.

This paper contributes to three lines of research. First, the measurement of ambiguity aversion is no longer the primary focus of this study's attention; rather, it focuses on whether subgroups take adaptive action in ambiguity scenarios, which is a significant departure from the previous ambiguity literature (Bajtelsmit, et al. 2015; Cabantous, 2007). This could be understood as an assessment of the decisions the general public makes in response to adjustments in public health policy. Second, we add to the literature on insurance WTP by focusing on how changes in external environmental effects interact with subjective attitudes. This is also consistent with the field's current focus trend, which is increasingly moving away from investigating people's social characteristics and toward studying how the environment plays a part (Zhang and Qian, 2018). Third, this study can advise governments (especially those in emerging economies) on how to determine the costs of modifying policies in light of shifting pandemic control claims. This paper highlights the impact of external ambiguity on individual economic behavior.

---

[1] The Joint Prevention and Control Mechanism of the State Council of China issued 20 Measures to Further Optimize Epidemic Prevention and Control. The Chinese document can be accessed at http://www.gov.cn/xinwen/2022-11/11/content_5726144.htm.

## 2. Experimental Procedure

The online experiment was carried out using Credamo software from e-Math Modeling Technology Co., Ltd. All participants signed a consent form and had their identifying information kept private. There were 672 participants in total, 384 of whom came from the COVID-19 high-risk communities identified by the Shijiazhuang CDC and 378 from low-risk areas. Subjects were informed they would receive the game proceeds plus a CNY 20 show-up fee. The game consisted of five distinct rounds, and before selecting how to respond to the ambiguity of public health policy in each round, subjects were reminded of the consequences of their choice to make sure they understood the experimental design.

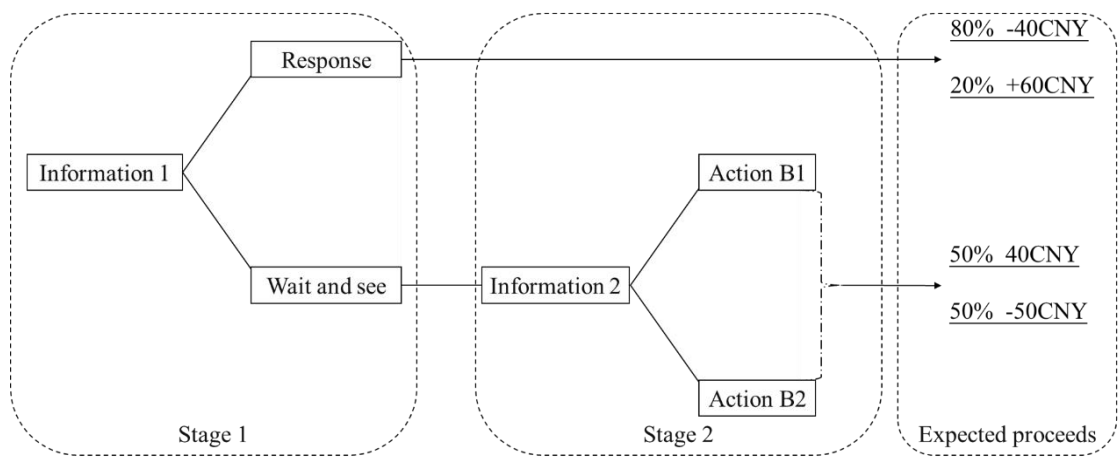

**Fig. 1.** The process of ambiguity game

Subjects were initially given an endowment of CNY 50 at the beginning of each round. Stages 1 and 2 were included in a single round (see **Fig. 1**).

In stage 1, subjects are given "Information 1", and depending on how they perceive the ambiguity, they can either select "Respond" or "Wait and see." When "Respond" is chosen, stage 2 is skipped, and the subject begins the ball picking game. Subjects select balls from "Box I", which contains eight red and two blue balls. If the red ball is picked, the payoff is -40 CNY; if the blue ball is picked, the payoff is 60 CNY, and the round is over. When subjects choose "Wait and see", they will move on to stage 2.

In stage 2, subjects are provided "information 2", which is complementary to the preceding information. Public health policy, to a lesser extent, remains ambiguous. Based on the new information, subjects can choose between "Action B1" and "Action B2". Whatever option subjects select in this phase, the ball will be drawn from "Box II", which includes 5 black balls and 5 yellow balls. When a black ball is selected, "Action B1" receives CNY -50 and "Action B2" receives CNY 40; when a yellow ball is selected, "Action B2" receives CNY -50 and "Action B1" receives CNY 40, and the round ends.

The game's design implies that while there is a significant likelihood of making a mistake at times of high public health policy ambiguity (i.e., when there is an early shift in policy), the cost of reacting is relatively modest. The subjects pick a card from Box III after five rounds. Each of the five cards in Box III represented one round. The final game proceeds are determined to be the round payoff indicated by the card the subject chose. The subjects were categorized based on how frequently they selected "Response" in Stage 1. The subjects are classified as sensitive objects in the context of ambiguity and given a value of 1 for the variable *Response* when the number of occurrences is greater than or equal to 3. In contrast, the variable *Response* is set to 0.

After the game, subjects completed a questionnaire on their WTP for a detailed commercial health insurance policy, as well as sociodemographic data. The terms are simplified from a commercial health insurance policy called *Yanzhao* Health Insurance that is offered in Shijiazhuang. The details of the single policy include but are not limited to a deductible of CNY 20,000, a CNY 3 million insurance amount, and a one-year insurance period. We also asked subjects to score their own health (in comparison to their peers) and their experience with income disruption due to epidemic control. **Table 1** provides a statistical summary of the variables, and **Table 2** provides a list of the variables' names and definitions.

**Table 1**

Summary statistics.

| Variables | Obs. | Mean | Std. Dev. | Min | Max |
|---|---|---|---|---|---|
| WTP | 762 | 0.703 | 0.444 | 0.050 | 1.663 |
| Respouse | 762 | 0.600 | 0.490 | 0 | 1 |
| Lrisk | 762 | 0.496 | 0.500 | 0 | 1 |
| Age45 | 762 | 0.255 | 0.436 | 0 | 1 |
| Male | 762 | 0.787 | 0.409 | 0 | 1 |
| Ehealth | 762 | 2.455 | 0.925 | 1 | 5 |
| Edu | 762 | 0.441 | 0.497 | 0 | 1 |
| Marriage | 762 | 0.845 | 0.362 | 0 | 1 |
| Income | 762 | 95.07 | 140.7 | 8 | 1900 |
| Asset | 762 | 1395 | 2097 | 10.8 | 18000 |
| Noninterruption | 762 | 0.899 | 0.302 | 0 | 1 |
| Community | 762 | 43.92 | 11.38 | 23 | 65 |
| District | 762 | 2.648 | 1.334 | 1 | 6 |

**Table 2**

Variable definitions.

| Variable name | Definition |
|---|---|
| WTP | Amount of subject's willingness to pay for the specified commercial health insurance (in CNY thousand). |
| Respouse | A dummy variable equals 1 if the subject's experimental result is marked as sensitive in the context of ambiguity and 0 otherwise. |
| Lrisk | A dummy variable equals 1 if the subject's community is a low-risk area and 0 otherwise. |
| Age45 | A dummy variable equals to 1 if the subject's age is between 45 and 60 years and 0 otherwise. |
| Male | A dummy variable equals 1 if the subject is male and 0 otherwise. |
| Ehealth | A categorical variable describing the subject's evaluation of his or her own health in comparison to the level of his or her peers 1 is the healthiest, and 5 is the worst. |
| Edu | A dummy variable equals 1 if the subject has a high school education or higher and 0 otherwise. |
| Marriage | A dummy variable equal to 1 if the subject is in marriage or cohabitation and 0 otherwise. |
| Income | Subjects' annual household income (in CNY thousand), |
| Aasset | Subjects' household assets (in CNY thousand), |
| Noninterruption | A dummy variable equals 1 if the subject has no income disruption experience and 0 otherwise. |
| Community | Number of the subject's community, |
| District | Number of the subject's administrative district, |

## 3. Statistical Methods

Many factors directly affect residents' responses to policy ambiguity because of the wide variation across communities. To deal with endogeneity, we employ the DID (difference-in-difference) concept (Atkin, 2016). The first differential is the difference in subjects' different response propensities in the experimental results, and the second differential is the difference in the subject's neighborhood's high or low risk (flagged by the CDC during the experiment). Residents in low-risk areas can move freely and are more likely to experience ambiguity as a result of the disparity between policy content and reality. The model is as follows:

$$Ln(WTP_i + 1) = \alpha + \beta_1 Response_i \times Lrisk_c + \gamma_a + \lambda_\hbar + X_i^{'}\varphi + \varepsilon_i \quad (1)$$

where $Ln(WTP_i + 1)$ is the natural logarithm of subject $i$'s WTP for insurance. The interaction term $Treat_{ic} = Response_i \times Lrisk_c$, our variable of interest, is assigned a value of 1 when the subject is sensitive to the ambiguity identified in the experiment and lives in a low-risk community and 0 otherwise. We include the *district* fixed effects ($\gamma_a$), which incorporate sociocultural and economic effects across regions. Furthermore, we control the *health evaluation* fixed effects ($\lambda_\hbar$) to control any unobservable factors in individual health levels that may be related to insurance expenditures. $X_i^{'}$ are the control variables for individual and household characteristics. $\varepsilon_i$ denotes the random error term. Standard errors are clustered at the community level.

## 4. Result

Column (1) of **Table 3** shows the OLS results without controlling for factors other than the interaction term. After controlling for variables and fixed effects, as shown in Column (2), the group sensitive to policy ambiguity's WTP for commercial health insurance is 12.2% ($P < 0.01$) higher than the baseline.

In Columns (3) and (4), we divide the groups based on whether income has been disrupted as a result of epidemic control. It is believed that going through this experience will significantly alter how people view epidemic response policies and their own coping strategies. After excluding a specific sample, the results in Column (3) can be interpreted as a robustness test. Column (4) presents the regression results for the group with this experience, which are significant at the 10% level; statistically, the estimated results of the interaction terms are substantially larger than the group without this experience.

**Table 3**

The impact of ambiguity on WTP.

|  | Dependent variable: $Ln(WTP_i + 1)$ | | | |
| --- | --- | --- | --- | --- |
|  | (1) | (2) | (3) | (4) |
| $Response_i \times Lrisk_c$ | 0.135*** | 0.122*** | 0.119*** | 0.260* |
|  | (0.029) | (0.028) | (0.032) | (0.100) |
| Control variables | N | Y | Y | Y |
| *Districts* FE | N | Y | Y | Y |
| *Health evaluation* FE | N | Y | Y | Y |
| Observations | 762 | 762 | 685 | 77 |
| $R^2$ | 0.378 | 0.41 | 0.393 | 0.698 |

*Notes:* Standard errors in parentheses. * p<0.1, **p<0.05, *** p<0.01.

## 5. Conclusion

This paper contributes to the study of WTP for insurance in the context of policy ambiguity. The results show that when public health policy ambiguity is revealed, the WTP for commercial health insurance is 12.2% higher than the baseline for those who actively respond accordingly to reduce losses (or actively engage in self-protection measures), and the increase in WTP is more pronounced for individuals with income disruptions. Our study provides new evidence that individuals may change their economic behavior in the short term in the face of ambiguous policies and may even eliminate charity hazards (Brunette, et al. 2013). Future research could explore ways to control policy ambiguity to create incentives for specific economic behaviors.